\pdfoutput=1

\documentclass[5p]{elsarticle2}

\usepackage{hyperref,color}
\usepackage{multirow}

\begin{document}

\begin{frontmatter}

\title{Modeling cognitive deficits following neurodegenerative diseases and traumatic brain injuries with deep convolutional neural networks}

\author{Bethany Lusch\corref{cor}}
\cortext[cor]{Corresponding author at: University of Washington, Lewis Hall \#202, Box 353925, Seattle, WA 98195-3925 USA}
\ead{herwaldt@uw.edu}

\author{Jake Weholt\corref{}}

\author{Pedro D. Maia\corref{}}

\author{J. Nathan Kutz\corref{}}

\address{Department of Applied Mathematics, University of Washington, United States}

\begin{abstract}
The accurate diagnosis and assessment of
neurodegenerative disease and traumatic brain injuries (TBI) remain open challenges to doctors and medical practitioners.  
Both cause cognitive and functional deficits due to the diffuse presence of focal axonal swellings (FAS), but it is difficult to deliver a prognosis due to our limited ability to assess damaged neurons at a cellular level {\em in vivo}. 
In this paper, we simulate the effects of neurodegenerative disease and TBI using convolutional neural networks (CNNs) as our model of cognition.  CNNs were originally inspired by neuroscience and the hierarchical layers of neurons used for processing input stimulus to the brain. 
We start with CNNs pre-trained to perform classification, then utilize biophysically relevant statistical data on FAS to damage the connections in a functionally relevant way. In order to improve the model, we incorporate the idea that brains operate under energy constraints by pruning the CNNs to be less over-engineered. Qualitatively, we demonstrate that damage to the connections leads to human-like mistakes. Our experiments also provide quantitative assessments of how accuracy is affected by various types and levels of damage. The deficit resulting from a fixed amount of damage greatly depends on which connections are randomly injured, providing intuition for why it is difficult to predict the impairments resulting from an injury. There is a large degree of subjectivity when it comes to interpreting cognitive deficits from dynamically evolving complex systems such as the human brain. However, we provide important insight and a quantitative framework for several disorders in which FAS are implicated, such as TBI, Alzheimer's, Parkinson's, and Multiple Sclerosis. 

\end{abstract}


\end{frontmatter}

\section{Introduction}
\begin{figure*}
\centering
\includegraphics[width=\textwidth]{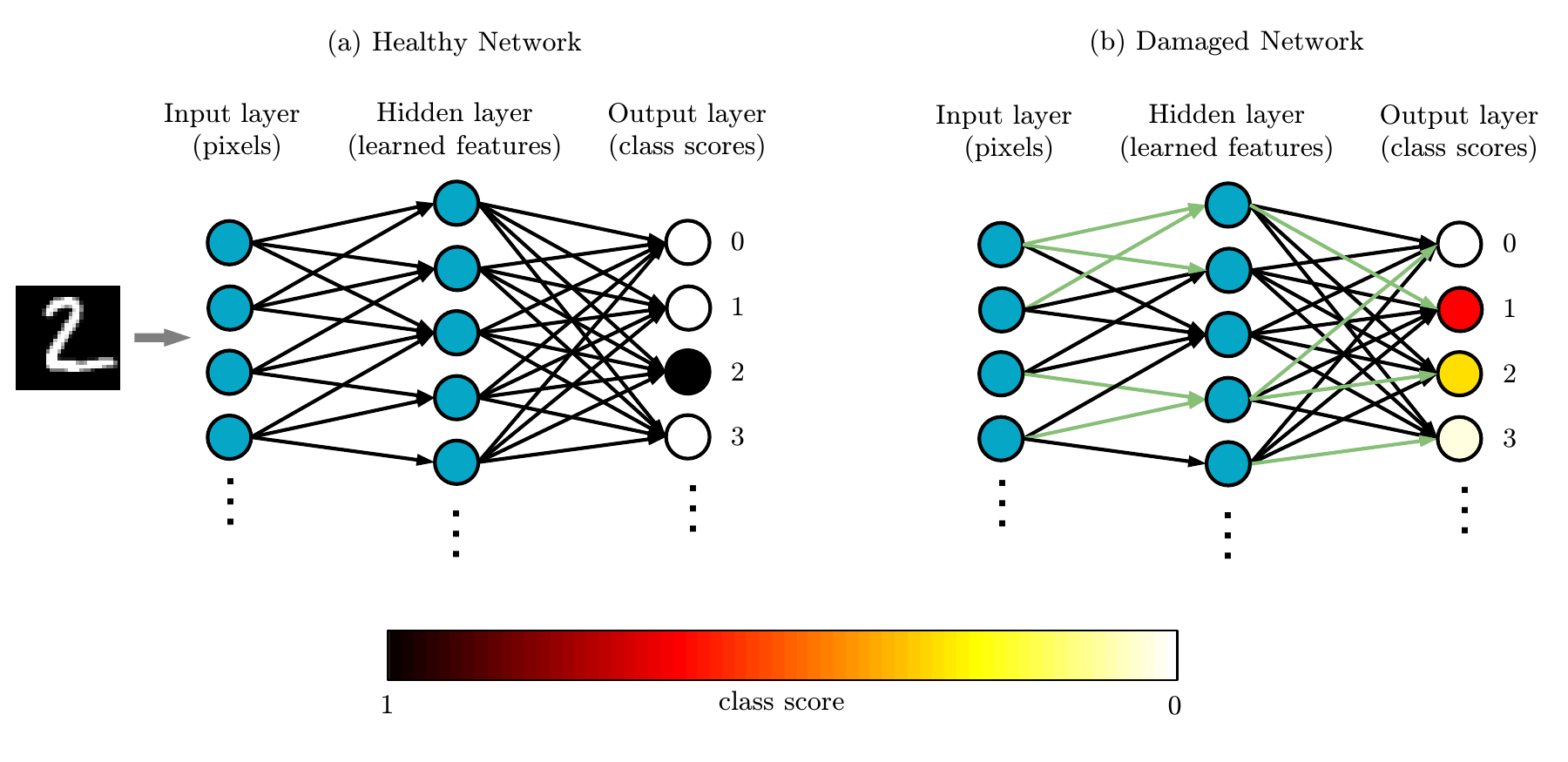}
\caption{Damaging a Convolutional Neural Network (CNN). (a) We start with a ``healthy'' CNN that accepts an image of a handwritten digit as an input and outputs scores for each possible digit, 0-9. We classify the image as the digit with the highest score. (b) We then damage the weights on the network in a biophysically-relevant way. In this figure, the healthy network correctly classifies the image as a 2, but the damaged network classifies it as a 1.}
\label{Overview}
\end{figure*}
The mathematical architecture of convolutional neural networks (CNNs) was originally inspired by the Nobel prize-winning work of Hubel and Wiesel on the primary visual cortex of cats~\cite{Hubel1962}.  Their seminal experiments were the
first to suggest that neurons in the visual system organize themselves in hierarchical layers of cells for processing visual stimulus.  The first quantitative model of the CNN, termed the Neocognitron by Fukushima in 1980~\cite{Fuku1980}, already displayed many of the characteristic features of today's deep CNNs, including a multi-layer structure, convolution, max pooling and nonlinear dynamical nodes. The connection between neuroscience and CNN theory, although clearly a conceptional abstraction~\cite{poggio2016}, has since been instrumental to improving quantitative models of how the brain integrates neuro-sensory information for stimulus classification and decision making.  Given that CNNs mimic many of the important cognitive features of the brain, we use it as a model for understanding how neurodegenerative diseases and traumatic brain injuries (TBI) can compromise an array of recognition tasks.  Specifically, by using well-established biophysical data on the statistics (distribution and size) of focal axonal swellings (FAS), which area among the primary symptoms of neurodegeneration and TBI, we evaluate the progress of impairments on a CNN-based model of cognition.  Our model provides quantitative metrics for understanding how cognitive deficits are accumulated as a function of FAS development, allowing for potentially new diagnostics for the evaluation of brain disorders due to neurodegenerative diseases and/or TBI.

Understanding how neurodegenerative diseases and TBI affect cognitive function remains a critically important challenge for societal mental health.  TBI alone is one of the major causes of disability and mortality worldwide, which in turn, dramatically jeopardizes society in several socioeconomic ways~\cite{Menon2015}.  Not only is it the signature injury of the wars in Afghanistan and Iraq~\cite{Jorge2012}, it is also the leading cause of death among young people~\cite{cdc}.  While many survive the events that induce TBI, persistent cognitive, psychiatric, and physiological dysfunction often follows from the mechanical impact (see Sec. 2).  Likewise, neurodegenerative diseases are responsible for an overwhelming variety of functional deficits, with common symptoms including memory loss or behavioral/cognitive impairments which are related to an inability to correctly process multi-modal information for decision-making tasks.  The majority of brain disorders have a complex cascade of pathological effects spanning multiple spatial scales: from cellular or network levels to tissues or entire brain areas.  Unfortunately, our limited ability to diagnose cerebral malfunctions {\em in vivo} cannot detect several anomalies that occur on smaller scales.  FAS, however, are ubiquitous to TBI and most leading and incurable disorders that dramatically affect signaling properties of neurons,  such as Multiple Sclerosis, Alzheimer's and Parkinson's diseases. 

Given the currently available wealth of data on FAS morphology from TBI studies and from almost every leading neurodegenerative disease, significant progress can be made towards understanding qualitatively how FAS impacts cognitive function.  In this work, we consider a set of deep CNN models 
as an \textit{abstraction} for functioning brains.  Our goal is to understand how the processing of input data (classification) is compromised as a function of increasing injury and/or disease progression.   Of course, it is obvious that the system's performance will be compromised as the CNN is injured, but the \textit{manner} in which the cognitive impairments arise is quite illustrative and informative, providing intuitively appealing results about how cognitive deficits can develop and evolve as a neurodegenerative disease progresses.  

Figure \ref{Overview} illustrates our approach. We begin with the original (healthy) CNN, which is trained to perform a classification task. In Figure \ref{Overview}, the specific task is to label a handwritten digit. We then expose the CNN to different injury protocols based upon biophysical observations of FAS statistics and morphological parameters. In particular, we use statistical distributions of FAS from a recent experiment consisting of TBI-induced damage in the visual cortex of rats \cite{Wang2011}.   To impose these injury statistics on the original CNN, we assume that each neuronal connection has a biophysically plausible probability to malfunction;  while mild axonal injury may simply weaken a connection, severe cases may break it permanently  (i.e., an axomtotomy occurs so that the connection strength goes to zero).  Ultimately, the severity of the injury and re-weighting of connections is also determined by biophysical data and the statistical distribution of the size of the FAS. We can then progressively monitor the 
deleterious effects of the injury on the functionality of the CNN, providing metrics for cognitive deficits that arise.

The paper is outlined as follows:  In Sec.~\ref{background} we provide key background material on the two primary fields integrated into this work: convolutional neural networks and neural disorders in which FAS are implicated. We describe our methodology in Sec.~\ref{method} and present results in Sec.~\ref{results}. We summarize our conclusions in Sec.~\ref{concl}. For full details, all MATLAB and Python codes used for this paper are available online at \url{github.com/BethanyL/damaged_cnns}.

\section{Background\label{background}}

\subsection{Convolutional Neural Networks}

Deep convolutional neural networks (DCNNs) are transforming almost every field of science involving big data.   The
success of the method has been enabled by two critical components:  (i) the continued growth
of computational power (e.g. GPU and networked computing), and (ii) exceptionally large labeled data sets capable of taking advantage of the full power of a multi-layer architecture.  Indeed, although the theoretical inception of CNNs has an almost four-decade history, the analysis~\cite{IN_Hinton} of the ImageNet data set~\cite{imageNET} in 2012 provided a watershed moment for CNNs and Deep Learning~\cite{LeCunreview}.  Prior to this data set, there were a number of data sets available with approximately tens of thousands of labeled images.  ImageNet provided over 15 million labeled, high-resolution images with over 22,000 categories.  DCNNs have since transformed the field of computer vision by dominating the performance metrics in almost every meaningful computer vision task intended for classification and identification (see, for example, the International Conference on Computer Vision 2015).

ImageNet has been a critically enabling data set for the evolution of the field.  However, CNNs were a topic of intensive research long before.   Indeed, they were highly successful in a wide range of applications and machine learning architectures.   By the early 1990s, neural networks were studied as standard textbook material~\cite{Bishop_CNN}, with the focus typically on a small number of layers.  Critical machine learning tasks such as principal component analysis (PCA) were shown to be intimately connected with networks which included back propagation~\cite{baldi,sanger1989optimal}.  Importantly, there were a number of critical innovations which established multilayer feedforward networks as a class of universal approximators.  Specifically, Hornik et al.~\cite{hornik1989multilayer} rigorously established that standard multilayer feedforward networks with as few as one hidden layer using arbitrary squashing functions were capable of approximating any Borel measurable function from one finite dimensional space to another to any desired degree of accuracy, provided sufficiently many hidden units were available. Thus, multilayer feedforward networks could be thought of as a class of universal approximators~\cite{hornik1989multilayer}.

The past five years have seen tremendous advances in the DCNN architecture.  Innovations have come from algorithmic
tricks and modifications that have led to significant performance gains in a variety of fields.  These innovations include pretraining~\cite{hinton2006fast,bengio2007greedy,erhan2010does}, dropout~\cite{srivastava2014dropout}, max pooling~\cite{IN_Hinton},
inception modules~\cite{szegedy2015going}, data augmentation (virtual examples)~\cite{niyogi1998incorporating}, batch normalization~\cite{ioffe2015batch} and/or residual learning~\cite{he2015deep}.  This is only a partial list of potential algorithmic innovations available for improving the performance of classification and labeling.  Our goal is not to provide a complete review of the DCNN field, but rather to highlight the continuing and rapid pace of progress in the field.  Integrating the state-of-the-art in DCNNs is the open source software called TensorFlow (tensorflow.org).  TensorFlow was originally developed by researchers and engineers working on the Google Brain Team within Google's Machine Intelligence research organization. The system is designed to facilitate research in machine learning and to make it quick and easy to transition from research prototype to production system.  TensorFlow has allowed for the test-bedding of new algorithmic structures in a reproducible and verifiable manner, which is a significant and important advancement in the field.
Indeed, the DCNN architecture used here relies on the TensorFlow architecture, helping us understand how
state-of-the-art DCNNs relate to cognitive abilities.

\subsection{Focal Axonal Swellings \label{FAS}}
\begin{figure}
\centering
\includegraphics[width=3.54in]{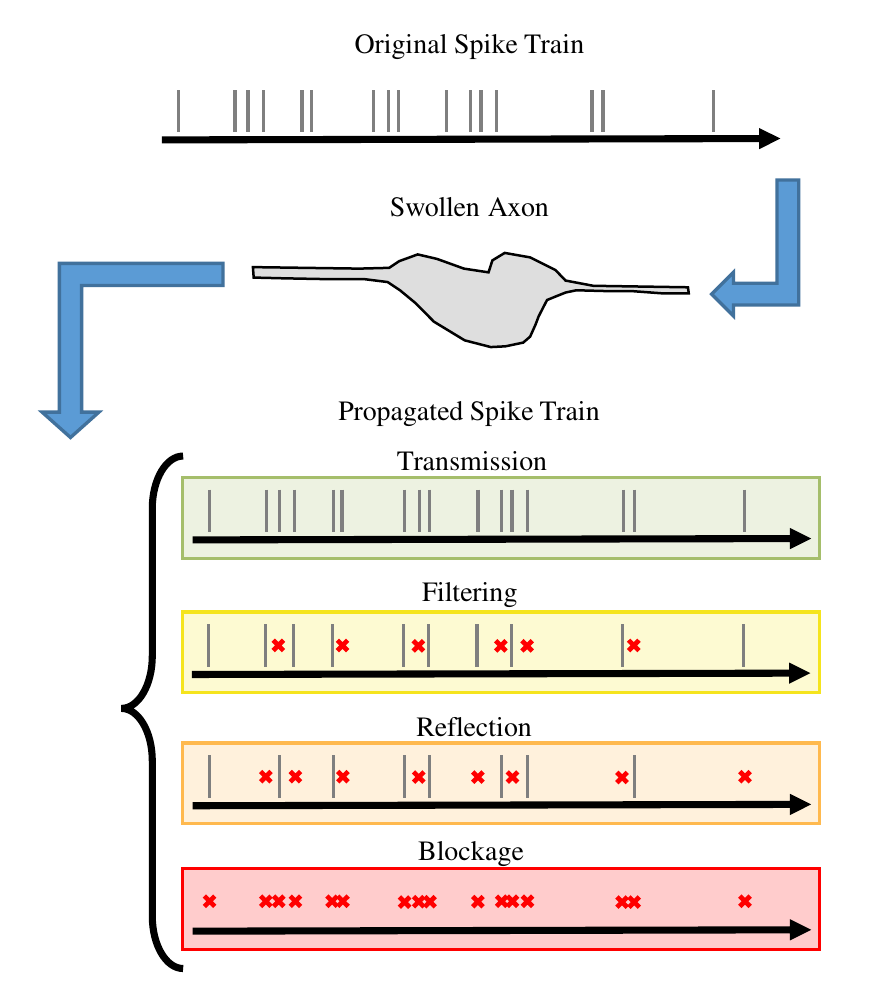}
\caption{Four Types of Damaged Axons. A spike train passes through a swollen axon. Depending on the way that the axon is swollen, there are four ways that the information can be transmitted. In transmission, the spike train is propagated correctly despite the damage. In filtering, the spike train goes through a low-pass filter. Regions of the spike train with high frequency are especially likely to lose spikes. In reflection, pairs of spikes combine and only half of the spikes are transmitted. In blockage, none of the spikes are transmitted.}
\label{Damage}
\end{figure}

\noindent Concussions and Traumatic Brain Injuries (TBI) are more than ever a concern for contact sport practitioners \citep{book_SI}, 
for veteran soldiers exposed to blast injuries \citep{Mau2016,Jorge2012}, and for society as a whole \citep{cdc,Menon2015,Roozenbeek2013}. 
In fact, TBI contributes to one-third of all injury-related deaths and is one of the major sources of functional 
\href{http://www.cdc.gov/traumaticbraininjury/data/}{impairments}. 
TBI pathologies affect several spatial scales \citep{Sharp2014}, but a ubiquitous development at the
neuronal microenvironment level is the presence of axonal injury \citep{Hill2016,Johnson2013,Skandsen2010}. 
As reviewed in \cite{Hill2016}, rapid axonal stretch injury triggers secondary axonal changes that can vary in extent and severity 
\citep{Edlow2016,Hanell2015,Henninger2016}, but most often culminate in Focal Axonal Swellings (FAS). 

FAS are monitored whenever possible in \textit{in-vitro} studies
\citep{Chen2009,Fayaz2000,Hellman2010,Hemphill2011,Hemphill2015,Magdesian2012,Morrison2011,Smith1999},  
in \textit{in-vivo} experiments \citep{Browne2011,Dikranian2008,Maxwell1997,Wang2011},  and in human patients 
\citep{Adams2011,Blumbergs1995,Christman1994,Grady1993,Jorge2012,Kinnunen2010,Povlishock2005}. 
In many cases, FAS critically affect the axonal morphology \citep{TangSchomer2012,TangSchomer2010}
and consequently, the information content encoded in spike trains propagating throughout them. 

Recent computational studies distinguished geometrical axonal enlargements that lead to minor changes in propagation from those that 
result in critical phenomena such as reflection or blockage of the original traveling pulse  \cite{Maia2014_1}, or filtering of action 
potentials \cite{Maia2014_2}. This led to a diagnostic toolbox that extracts meaningful geometrical parameters from sequential images 
of injured axon segments \cite{Maia2015}. These algorithms provide a principled approach to deal with imaging distortions caused by 
experimental artifacts in order to extract the cross-section of an axon by detecting local symmetries, turning points and turning regions. 
More importantly, they provide the first description of biologically plausible injurious effects due to FAS that can be incorporated into neuronal 
network simulations. Figure \ref{Damage} reviews these different effects.  In the transmission regime, the spike train propagates through the FAS 
without significant modifications. In the filtering regime, pulses that are too close to each other get deleted by a mechanism resembling a pile-up collision 
\cite{Maia2014_2}. As the FAS geometrical parameters worsen, a single spike will split into two components, one propagating forward and the 
other propagating backward. The reflected, back-propagating pulse will collide with the next spike in the train and they will mutually annihilate each 
other. Thus, the reflection regime effectively halves the firing rate of the neuron. Finally, in the worst-case scenario, the FAS will block all spikes 
and transmit no information 
whatsoever.  

In what follows, we will introduce FAS in a biologically plausible way to a few examples of deep-learning convolutional neural networks 
and evaluate the extent to which cognitive deficits develop. 

\section{Materials and Methods\label{method}}
\small{
\begin{table*}[!t]
\renewcommand{\arraystretch}{1.3}
\caption{Summary of CNNs Used}
\label{TableCNN}
\centering
\begin{tabular}{|c|c|c|c|c|c|}
\hline
\multirow{2}{*}{Task} & \multirow{2}{*}{Training Set} & \# conv. &  \# fully connected & \# weights \\
& & layers & layers & to damage\\
\hline
handwritten digit classification & MNIST: 55K & 2 & 1 & 83K\\
\hline
\multirow{2}{*}{object classification} & ImageNet, ILSVRC & \multirow{2}{*}{5} & \multirow{2}{*}{3} & \multirow{2}{*}{61M} \\
&  2012: 1.2M & & & \\
\hline
face verification & VGG-Face: 2.6M & 13 & 2 & 134M\\
\hline
\end{tabular}
\end{table*}
}
\normalsize
\subsection{CNN training, calibration, and performance}
We simulate the development of FAS damage in three different convolutional neural networks. 
Each network has its own properties and was trained with different data sets for separate tasks  (see Table \ref{TableCNN}).

First, we consider a network trained on the MNIST dataset \cite{mnist}, which is composed of images of handwritten digits. We train the network to classify each image as a digit from 0 to 9. It could be used, for example, by a post office machine to read zip codes from envelopes. The training data set consists of a series of black and white images that are 28 by 28 pixels. We use the TensorFlow framework \cite{TensorFlow} to train a CNN with two convolutional layers and a fully connected layer, as advised by a TensorFlow tutorial. We use a subset of the standard MNIST test set for our testing purposes so that our set contains the same number of examples for each digit.  In particular, we choose the first 852 images for each digit. 
Our trained network has an accuracy of 98.74\% on this test set. 

Next, we use a network trained on the ImageNet data set to classify images from the ILSVRC 2012 challenge as one of one thousand objects \cite{ILSVRC15}. 
The CNN-F network was pre-trained by the Visual Geometry Group at Oxford \cite{ImageNetNetwork} and made available through the MatConvNet Matlab Toolbox \cite{MatConvNet}, where it is referred to as imagenet-vgg-f. 
The network contains five convolutional layers and three fully connected layers. For our experiments, we use a subset of the data with two examples randomly chosen from each class. The network is 54.6\% accurate on this test set. 

The third network that we use was trained to classify faces as one of one thousand people. However, if you remove the last classification layer and normalize the output vector, the network can instead be used to create feature vectors for face verification. If the Euclidean norm of the difference between the feature vectors for two images is under a threshold $\tau$, the pair of images is classified as being the same person. This network was also trained by the Visual Geometry Group \cite{FaceNetwork} and made available through the MatConvNet toolbox \cite{MatConvNet}, where it is called vgg-face. For our experiments, we randomly choose five pictures each of fifty randomly chosen celebrities from the Labeled Faces in the Wild (LFW) data set \cite{LFW}. We also need to choose a threshold $\tau$. We choose $\tau = 1.2$ based on Linear Discriminant Analysis on a training set of 5700 examples from the LFW data set. Each of the 250 images in our test set is then compared to the four other images of the same person and four images of other people. We thus test one thousand pairs of images, half of which are of the same person and half of which are not.  The network is 81.6\% accurate with $\tau = 1.2$ on our test set of one thousand pairs of images. 

\subsection{Network impairments following FAS injuries}
To simulate the effects of traumatic brain injury on a CNN, we randomly ``damage'' $p$ percent 
of the weights in the convolutional and fully connected layers. For consistency with the TBI  analogy,
we only target the connections between neurons and not bias weights. Note that these CNNs are designed to use the same weights for multiple connections. Thus, damaging $p$ percent of the weights is not equivalent to damaging $p$ percent of the connections.  For simplicity, we first assume that
all axonal injuries lead to the total blockage of spikes, which effectively sets $p$ percent of weights to zero. 
We consider damage examples for each one of the previously described networks to develop  
intuition about possible functional impairments.

\begin{figure}
\centering
\includegraphics[width=3.54in]{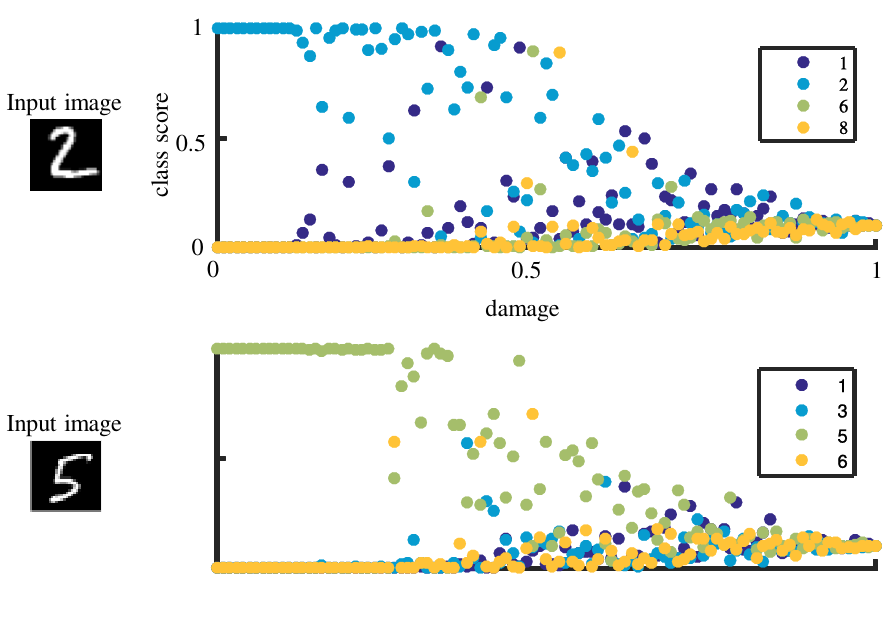}
\caption{Change in Class Scores on Damaged Networks. This CNN accepts an image of a handwritten digit as an input and outputs scores for each possible digit, 0-9. In these two examples, the original network correctly and confidently classifies the digit. As we increase the damage level, confidence drops and the classes eventually become confused. For high levels of damage, all classes have similar scores. }
\label{ExampleTwos}
\end{figure}

In Figure \ref{ExampleTwos}, we choose a handwritten ``2'' as the input to the MNIST network. The network assigns a score to each of the ten possible digits and then classifies the image as the digit with the highest score. The original network gives scores of $.999987$ to 2, $.000013$ to 1, and $0$ to the rest of the digits and thus correctly classifies the image as a 2. We then randomly damage the network in 100 separate experiments, setting $p = .01, .02, \dots, 1$. Since we are simulating TBI, the damage happens all at once and is not accumulated across experiments. Thus, the set of damaged neurons when $p = .01$ may have little overlap with the targeted neurons in the $p = .02$ case. At around $12\%$ damage, the network becomes noticeably less confident, but still correct. The network makes its first mistake at $30\%$ damage by labeling the image as a 1. At higher levels of damage, it frequently confuses classes 1 and 2. After $90\%$ damage, the ordering of the class scores continues to change, but their values become quite similar. We see an analogous pattern in the second part of Figure \ref{ExampleTwos} where we input a ``5'' as an example, although the damaged networks make fewer mistakes. 

\begin{figure*}
\centering
\includegraphics[width=1.0\textwidth]{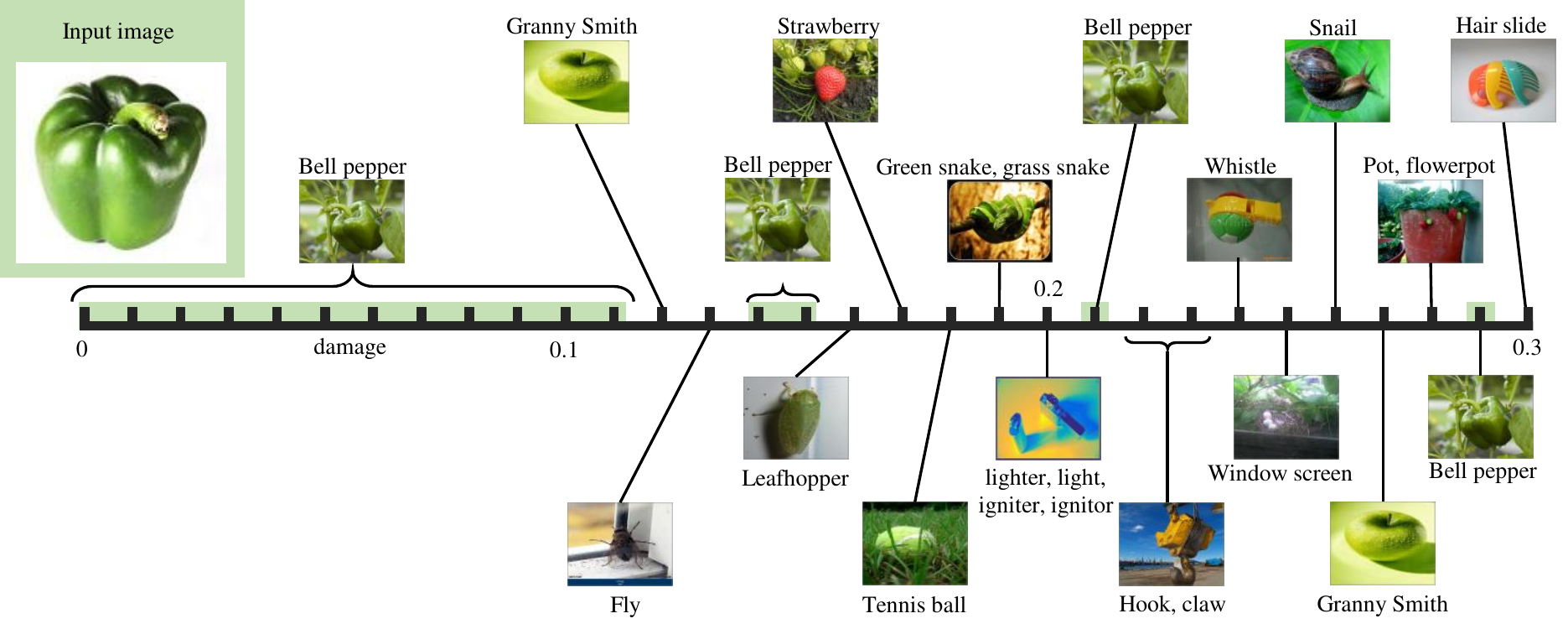}
\caption{Classification Mistakes as Damage Increases, Example 1. We start with a healthy network trained to classify images. The original network correctly classifies this image as a green pepper, but with enough damage, the network makes mistakes. For moderate amounts of damage, the wrong classifications make some intuitive sense.}
\label{Examples}
\end{figure*}

Next, we consider two examples from the image classification problem. 
In Figure \ref{Examples}, we use an image of a green bell pepper as the input to the ImageNet network. Here the network assigns a score to each of one thousand possible classes before classifying the image as the class with the highest score. 
We visualize how the classification changes as the damage increases $(p = 0, 0.01, 0.02, \dots, 0.3)$. The network makes its first mistake at $12\%$ damage but sometimes returns to classifying the image as a bell pepper. Some of the mistakes are relatively sensible, such as ``granny smith'' or ``tennis ball,'' and share similar colors, 
textures and/or shapes with the original image. Some of the later mistakes seem less understandable, such as ``hair slide.'' Note that this network was trained on about 1.2 million images of the one thousand classes, encompassing a wide range of examples for each class. For illustration purposes in this figure, we show an example image from the test set for each class. However, the input image is downloaded from Flickr \cite{pepper}.

\begin{figure*}
\centering
\includegraphics[width=1.0\textwidth]{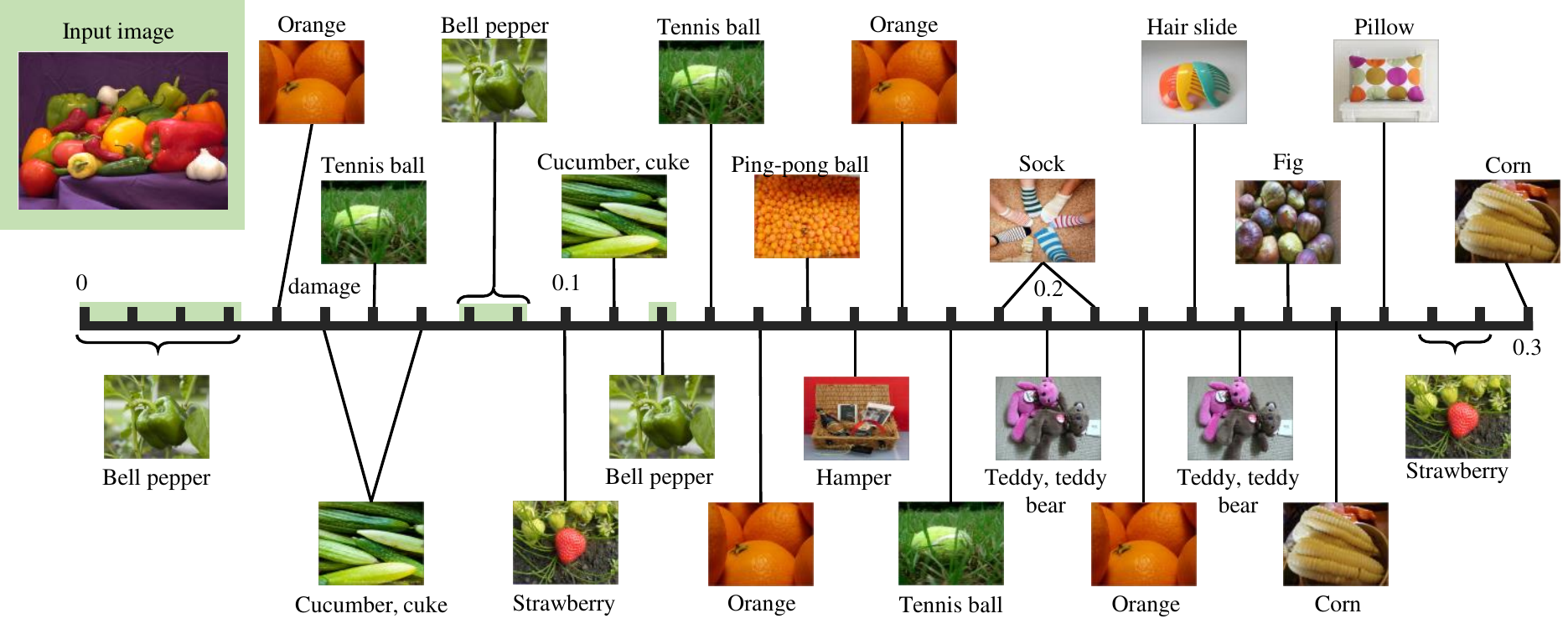}
\caption{Classification Mistakes as Damage Increases, Example 2. We increase the difficulty by using an image of a group of vegetables, primarily bell peppers. The network does not maintain the ``bell pepper'' classification as long, but the early mistakes are also produce or also round items.}
\label{ExamplesPepperGroup}
\end{figure*}

In Figure \ref{ExamplesPepperGroup}, we give a more difficult input image to the ImageNet network---a group of vegetables composed predominately of peppers with a variety of colors but also containing garlic. This image accompanies the Image Processing Toolbox for Matlab as peppers.png and is used as a demo for this network in the MatConvNet Toolbox. The network successfully chooses the bell pepper class among one thousand possibilities, but it is not as robust to injury 
as the one in the previous, easier example. 
Misclassifications begin at 4\% injury. Again, some errors are reasonable, such as 
a ``cucumber'' or ``orange'' (which is not that different from an orange-colored pepper). Others are quite surprising, 
such as ``socks''  or ``teddy bear''. 

\begin{figure}
\centering
\includegraphics[width=3.54in]{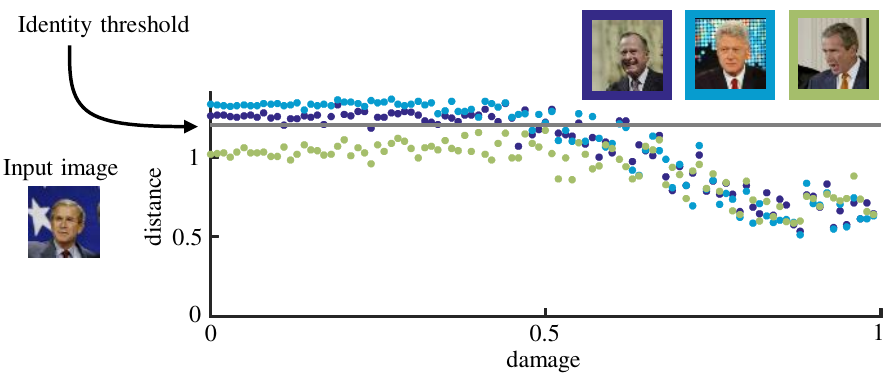}
\caption{Change in Distance Between Images. This network outputs a feature vector for each image and can be used to find the distance between two images. If the distance is below our threshold $\tau$, the pair is labeled as being of the same person. The network originally correctly identifies the second image of George W. Bush as being the same person while labeling the images of George H. W. Bush and Bill Clinton as being different people. After sufficient damage, the distances between the images all shrink and it is not possible to determine whether or not a pair of images are of the same person.}
\label{ExamplePrez}
\end{figure}

In Figure \ref{ExamplePrez}, we show analogous deficits for the facial recognition network (VGG-Face). We input an image of George W. Bush to the network and have it compare the image with three other pictures---one of his father George H. W. Bush, one of Bill Clinton, and another one of himself.  All images come from the LFW data set. The healthy network correctly identifies that both pictures of George W. Bush are of the same person and that the pictures of his father and Bill Clinton are of different people. We also see that George W. Bush is closer to his father than to Bill Clinton. As the damage increases, the network occasionally classifies George H. W. Bush as being the same as his son and, eventually, cannot even distinguish Bill Clinton. After about $70\%$ damage, all images start looking alike, and the network continuously exchanges the ordering of the distances as the damage increases. This pattern continues in the broader experiments and, with enough damage, all pairs of images are labeled as being of the same person. Note that adjusting the threshold as damage increases would not improve the accuracy since the second picture of George W. Bush does not remain closer than the pictures of other people. 

\section{Results\label{results}}
\noindent In this section, we move from qualitative descriptions of single network errors to a more broad, 
statistical account of mistakes within the test sets. We also consider a few variations of FAS 
injury protocols, network settings, and their dynamics to model biologically relevant phenomena such as aging and the development of neurodegenerative effects across CNNs.
\subsection{Overall network impairments} 
\begin{figure}
\centering
\includegraphics[width=3.54in]{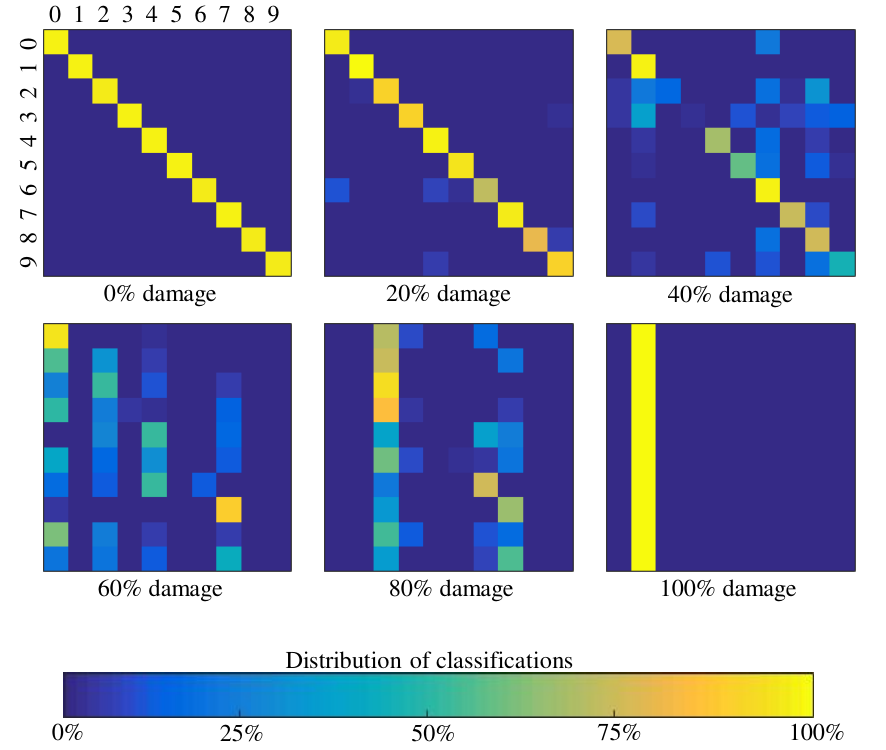}
\caption{Confusion matrices as damage increases. We depict the classification results of the handwritten digit classification network for varying amounts of damage. If the images are perfectly classified, only the diagonal is non-zero. As the damage increases, most images are mapped to the same few digits. Eventually, all images are classified as a ``1.''}
\label{ConfusionMatrices}
\end{figure}

In Figure \ref{ConfusionMatrices}, we return to the MNIST handwritten digit classification task and plot confusion matrices $M_{i,j}$ for the ten digits as the damage percentage $p$ increases.  At $p = 0\%$, the network has $98.75\%$ accuracy, so the matrix is concentrated on the diagonal ($i = j$). At $p = 20\%$, we begin to see substantial errors ($i \neq j$), especially by over-classifying digits 0, 4, and 9. As the damage increases, the confusion matrices become even more distributed, but the types of errors change. For example, at $40\%$ damage, some especially common labels are 1 and 6, while at $60\%$ damage, the disproportionately common labels become 0, 2, 4, and 7. However, recall that in our TBI analogy, the damage is not accumulated---in each experiment, we return to the original network and randomly choose a new set of weights to damage.  At $p = 100\%$, there is no randomness; all weights are set to 0 and all images are labeled as a 1. Overall, confusion matrices provide a straightforward visualization for misclassification within the CNN 
data set that could be advantageous for diagnosing cognitive deficits. 

\begin{figure}
\centering
\includegraphics[width=3.54in]{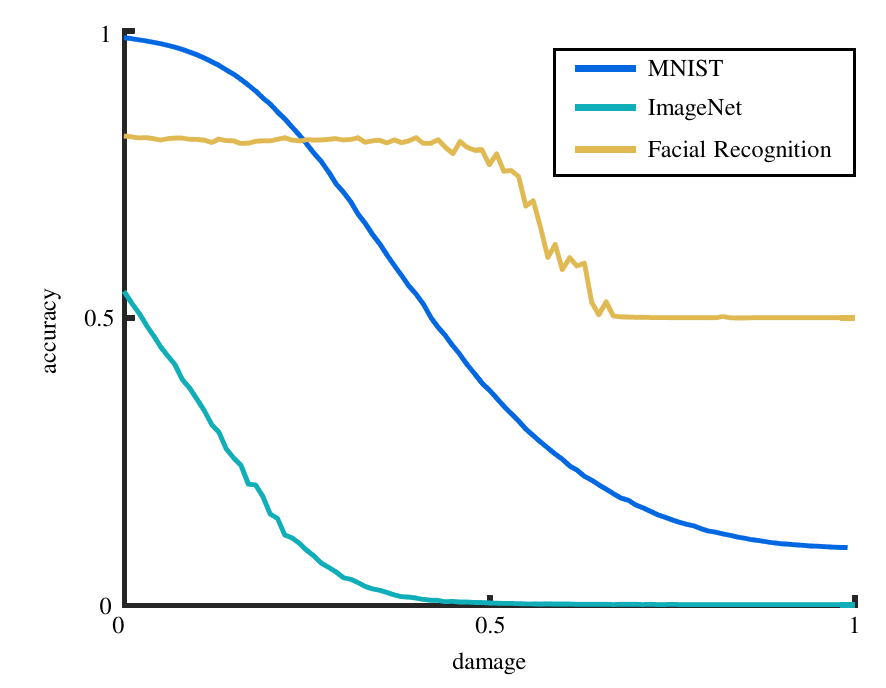}
\caption{Accuracy decay as damage increases. We randomly damage edges of the network by setting their weights to zero. We plot the percentage of edges that are damaged against the average accuracy of the network for three problems. We see that damage initially has little effect, but then there's a steep drop off until the accuracy levels off around the level of random guessing.} 
\label{TBI}
\end{figure}

In Figure \ref{TBI}, we summarize our TBI experiments for (i) the MNIST network, (ii) the
ImageNet network, and (iii) the Facial Recognition network. All targeted neurons are assumed to
malfunction the same way, fully blocking the signal transmission to their neighbors. For each 
damage level $p$ (\%), we average the accuracy across all random trials on that network. All three curves have a sigmoidal shape of the form \[
y = a\left(1 - \frac{1}{1+\exp(b(x+c))}\right) + \frac{1}{n},
\]
where $n$ is the number of classes (see Table \ref{sigmoid}). Again, there is no randomness when $p = 0\%$ or $100\%$. At $p = 100\%$, all weights are set 
to 0 and all examples are placed in the same class. Therefore, the accuracy of the network decays 
asymptotically to $1/n$. Note that the MNIST network and the 
ImageNet network have qualitatively similar trends, and display some accuracy deficit even at low
injury levels. On the other hand, the Facial Recognition network is able to maintain its original accuracy level
past $p = 50\%$ before decaying abruptly. 

\small{
\begin{table}[!t]
\renewcommand{\arraystretch}{1.3}
\caption{Sigmoid Parameters for Accuracy Drop-off Curves}
\label{sigmoid}
\centering
\begin{tabular}{|l|c|c|c|}
\hline
 & a & b & c\\
\hline
Figure \ref{TBI}, MNIST & 0.94 & -7.9 & -0.38 \\
\hline
Figure \ref{TBI}, ImageNet & 0.64 & -13 & -0.12 \\
\hline
Figure \ref{TBI}, Face Verification & 0.31 & -30 & -0.58 \\
\hline
Figure \ref{Sparsified}, combination & 1.2 & -2.2 & -0.42 \\
\hline
Figure \ref{Sparsified}, blockage & 0.98 & -6.3 & -0.25 \\
\hline
\end{tabular}
\end{table}
}
\normalsize

\subsection{Relevance of connections and biological constraints}
In all three examples of network dysfunction, there is a considerable amount of variability 
across trials even for the same injury levels. We found that the deficits greatly 
depend on \textit{which} weights were randomly selected. In other words, neuronal 
connections in CNNs do not contribute equally to a task, and damaging weights with 
large magnitude typically impacts the accuracy more than targeting weaker links---although 
magnitude alone cannot explain all cases. 

\begin{figure}
\centering
\includegraphics[width=3.54in]{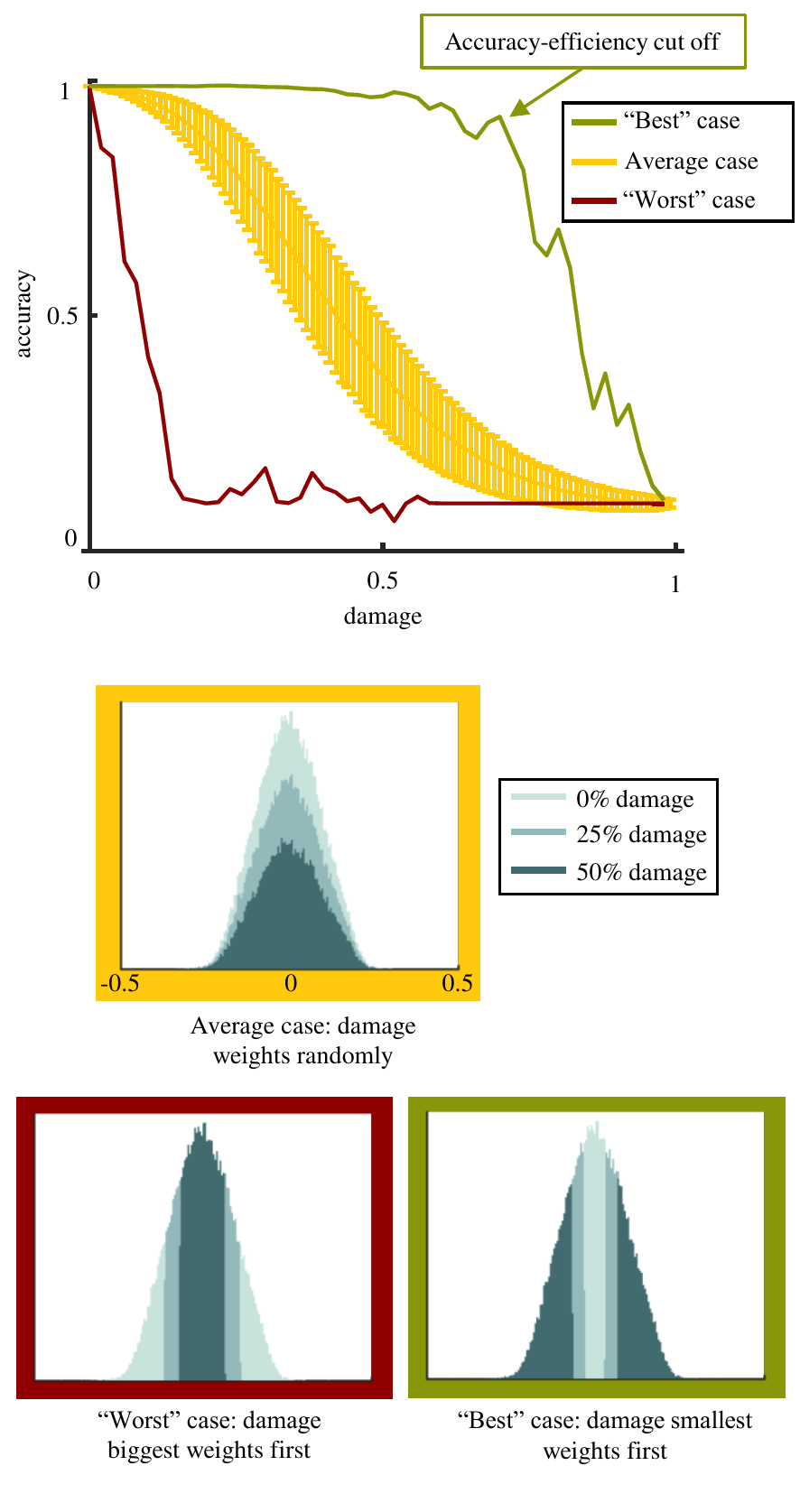}
\caption{Range of possible outcomes. The change in accuracy as weights are damaged varies depending on which weights were randomly chosen. In blue, we plot the average accuracy plus error bars for each level of damage. We also add curves in teal and yellow for approximations of best and worst-case accuracies, respectively. The approximate worst-case was found by damaging the weights in decreasing order of their absolute value. Similarly, the approximate best-case was found by damaging the weights in increasing order. We give a visualization in terms of a histogram of what it means to damage the weights in a random order (``average case''), in decreasing order (``worst case''), and in increasing order (``best case''). The yellow ``best case'' provides an accuracy-efficiency trade off. We choose a turning point in the curve: if we remove the smallest $69.4\%$ of the weights, the accuracy only decreases from $98.74\%$ to $91.47\%$.}
\label{RelImportance}
\end{figure}

We illustrate some of these issues in Figure \ref{RelImportance} for the the MNIST network. We repeat the average decay in accuracy in blue but add error bars. We found that roughly the worst case is to damage the weights in decreasing order of magnitude instead of randomly. The resulting steep accuracy drop off is plotted in teal. Conversely, the approximate best case is to damage the weights in increasing order of magnitude (plotted in gold). These three damage strategies are visualized in terms of their effect on the distribution of weights. The purple histogram displays the distribution of the original weights. In general, we randomly choose weights to damage, so the effect is distributed across the distribution of weights. However, damaging the weights in order of decreasing magnitude is equivalent to progressively removing the tails of the histogram, and choosing 
the weights in increasing order of magnitude is equivalent to removing the middle of the histogram. 
These experiments may provide intuition into why the outcomes of TBI are so difficult to predict. 
We hypothesize that randomness in the location of FAS could explain, for instance, why two 
soldiers near the same explosion site may develop significantly different post-traumatic outcomes. 

One of the most striking differences between artificial CNNs and biological neuronal networks is that the latter must operate under geometric, biophysical and energy constraints. As reviewed in 
\cite{Laughlin2003}, brains have evolved to operate efficiently since economy and proficiency are 
guiding principles in physiology. In fact, nervous systems are a major drain on an animal's energy 
budget and many aspects of the brain's anatomy seem to limit wiring costs \cite{Chklovskii2004,Kaiser2006,Sporns2011}. Brain networks can therefore be said to negotiate
an economical trade-off between minimizing inter-neuron connection cost \& maximizing 
topological value and capacity for information processing. See Bullmore and Sporns 
\cite{Bullmore2012} for a recent review on the topic. 

The MNIST network could be more biophysically plausible if it was not as over-engineered.
With its original topology and settings, the CNN becomes artificially resistant to damage.
In what follows, we will first \textit{sparsify} the CNN by picking a point on the accuracy-efficiency 
trade off curve (see Figure \ref{RelImportance}). There are multiple ways to choose the best trade-off point. 
A reasonable choice is to remove the weakest $69.4\%$ of the links, which decreases the accuracy 
from $98.74\%$ to $91.47\%$. 

\subsection{Different types of FAS dysfunctions}
As described in Section \ref{FAS}, Focal Axonal Swellings (FAS) affect spike trains in four qualitatively different
regimes: transmission, filtering, reflection, and blockage. So far we have only considered the worst case, blockage, which we model by setting a weight to zero. Now we also consider the other types of neuronal malfunctions. We model transmission as not damaging a weight, reflection as halving a weight, and filtering as applying a low-pass filter on each weight. We choose an example filtering function of $f(x) = -.2774 x^2 + .9094 x - .0192$ plus Gaussian noise $\sim N(0,0.05)$ by fitting a confusion matrix from experimental results \cite{Maia2014_2}. 
We believe that these additions to CNNs contain, in a tractable way, the key features of the 
jeopardizing effects caused by FAS described in \cite{Maia2014_1,Maia2014_2,Maia2015}.

\begin{figure}
\centering
\includegraphics[width=3.54in]{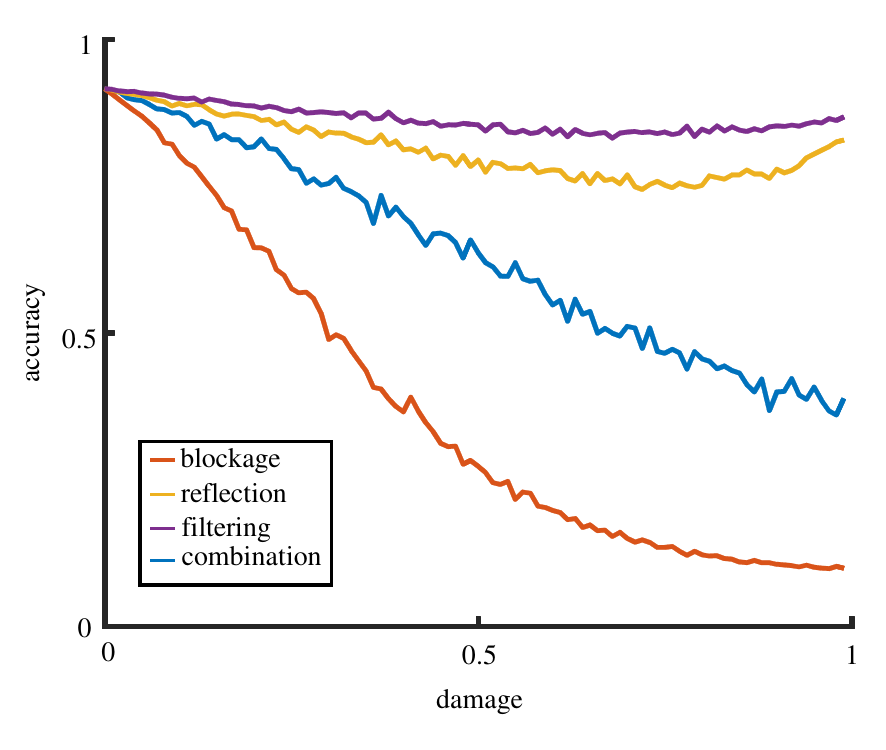}
\caption{Comparing types of damage. In these experiments, we begin with a ``sparsified'' network with the smallest $69.4\%$ of the weights removed. Then we compare the types of FAS (blockage, reflection, and filtering) and a combination of all types based on experimental evidence. As expected, blockage causes the most damage, and reflection is a strong form of filtering.}
\label{Sparsified}
\end{figure}

Recent experimental results provide detailed morphological descriptions of the FAS developing
after traumatic brain injuries. Wang et al \cite{Wang2011} damaged the optic nerve of adult rats
with a central fluid percussions injury. The optic nerve is a relatively organized bundle of axons and 
allowed for monitoring of FAS development 12h, 24h and 48h after the impact. They divided the nerve
in 12 serial grids and reported for each of them the number of axonal swellings per unit area, the total 
area of axonal swellings, and the individual sizes of swellings. It is possible to infer the fraction 
of FAS in each dysfunctional regime from these statistical distributions \cite{PedroThesis}. Based on these results, we conduct numerical experiments with 30\% blockage, 45\% reflection, 20\% filtering, and 5\% transmission. In Figure \ref{Sparsified}, we show results for the sparsified MNIST 
network, comparing its average accuracy for heterogeneous and homogeneous FAS distributions. 
As expected, the worst deficits occurred when all of the swellings were in the blockage regime. The best case is when all of the FAS are in the filtering regime, closely followed by the related reflection case. When we combine these regimes (30\% blockage, 45\% reflection, 20\% filtering, and 5\% transmission), the accuracy is between these more extreme cases. Although the accuracy decreases more moderately, it can still be fit with a sigmoid function (see Table \ref{sigmoid}).

\subsection{Aging and Neurodegenerative Diseases}

\noindent Alzheimer's Disease (AD) is the most commonly found type of dementia, which is an 
umbrella term for a variety of brain disorders and pathologies \citep{Jorm1998}. Aging is 
the single greatest risk factor for AD \citep{Patterson2015}, and most public health systems 
across the developed world are expected to face huge challenges due to the growing elderly 
population~\citep{Qiu2009}.
Recent estimates suggest that more than 5.2 million people have AD in the United States alone and that 
a new case occurs every 68 seconds  \citep{Thies2013}. The most typical symptom of the disease 
is an increasing difficulty in recalling new information, although it sometimes occurs in conjunction 
with challenges in completing familiar tasks, confusion with time or place, and trouble 
understanding visual images and spatial relationships.  
W. Thies and L. Bleiler \cite{Thies2013} report that in many cases,  AD diagnostics are accompanied by cognitive tests, since 
individuals with mild cognitive impairments have changes in thinking abilities that are noticeable to
family members and friends. We believe, however, that there is still a large degree of subjectivity 
when it comes to interpreting cognitive deficits from dynamically evolving complex systems such 
as the human brain. Thus, simulations with convolutional neural networks that incorporate biophysically 
plausible neural malfunctions may provide a window of opportunity to better diagnose, for instance, 
confusion in visual image classification. 

Focal axonal swelling pathologies are present in AD~\citep{Adalbert2009,Daianu2016,Krstic2012,Tsai2004} and in other neurodegenerative diseases
such as Parkinson's disease~\citep{Tagliaferro2016,Louis2009,Galvin1999},  Multiple Sclerosis~\citep{Friese2014,Nikic2011,Trapp2008}, and others~\citep{Herwerth2016,Karlsson2016,Laukka2016,Lauria2003}.
In many cases, FAS arise by the agglomeration of specific proteins over time \citep{Coleman2005,Millecamps2013}, 
and again, the computational modeling of focal axonal swellings and their effects to 
spike propagation from \cite{Maia2015} provide a platform to investigate network dysfunction.

\begin{figure}
\centering
\includegraphics[width=3.54in]{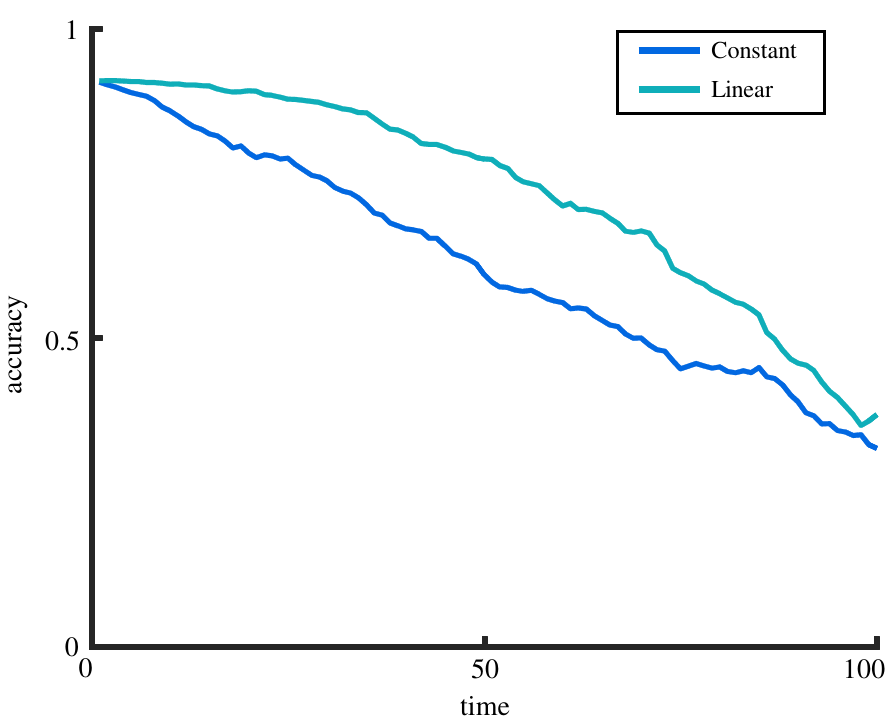}
\caption{Accumulating damage over time. In aging or neurodegenerative disease, damage to axons is accumulated over time, in contrast to a one-time injury. We compare the accuracy curves for a constant number of connections damaged for each time step to the case where the number of connections damaged increases with time. When damage increases over time, the initial loss in accuracy is slow and the later loss is faster.}
\label{Aging}
\end{figure}

In all of the previous experiments, we simulated TBI by abruptly applying axonal injuries. Here we instead simulate aging and its neurodegenerative effects by gradually accumulating random damage.  We continue to use the sparsified network and the heterogeneous FAS distribution.  Figure \ref{Aging} shows that  if damage is applied at a constant rate (targeting $1\%$ of the connections at each step), the results will look similar to a sequence of TBI experiments with $p = .01, .02, \dots$ (Figure \ref{Sparsified}, in dark blue) except that each trial will have a smoother trajectory. This is translationally relevant since traumatic brain injuries dramatically increase the risk 
of dementia later in life \cite{Barnes2014,Johnson2010,Johnson2012,Lobue2016}.
Perhaps a more plausible and biophysically relevant case occurs when the FAS accumulation rate 
increases linearly with time (in cyan). There, the young brain accumulates very little damage, but the older brain rapidly acquires new swellings. 

\section{Conclusions\label{concl}}
Assessing levels of cognitive deficits in patients is largely a subjective task, with indicators such as whether or not the patient and those close to them have noticed difficulties with memory. There are some tools available, including the Mini-Mental State Examination (MMSE).  The MMSE assigns a score after testing performance on a brief series of tasks such as identifying objects and following written instructions. This score can be used to quantitatively track changes in a person's cognitive function. Similarly, in this paper, we calculate the change in accuracy on related tasks, such as reading handwritten numbers and labeling objects. Since we can conduct extensive experiments with any level of injury, we believe that simulating FAS on our model of cognition can lead to insight into the complex processes underlying TBI and neurodegeneration.

Non-invasive diagnostic tools cannot detect anomalies {\em in vivo} such as FAS that occur at the cellular level. In fact, this has motivated a large body of {\em in vitro} experiments to replicate these injuries in a controlled setting \cite{Chen2009,Fayaz2000,Hellman2010,Hemphill2011,Hemphill2015,Magdesian2012,Morrison2011,Smith1999}. However, in the latter case, the cognitive effects of these injuries cannot be assessed. Simulations provide an opportunity to connect understanding of FAS to measures of cognitive performance. 

Both CNNs and brains operate somewhere on an accuracy-efficiency trade-off curve. However, it can be argued that brains are more highly constrained than CNNs due to the high energy costs of maintaining the nervous systems \cite{Laughlin2003}. In contrast, many CNNs are trained with a high focus on small gains in accuracy, especially those trained for competitions such as ImageNet. In addition, all three of the CNNs studied here utilized dropout, which encourages redundancy in the weights \cite{srivastava2014dropout}. A key step in our methodology was to prune the CNNs to be less over-engineered and thus more biologically plausible. Remarkably, the networks performed very well even if many weak connections were removed. 

Simulations of FAS damage to our CNN model of cognition result in interpretable and human-like mistakes, such as confusing a handwritten ``5'' with a ``6'' (Figure \ref{ExampleTwos}), labeling peppers as an apple or a cucumber (Figures \ref{Examples} and \ref{ExamplesPepperGroup}) and confusing George W. Bush with his father (Figure \ref{ExamplePrez}). We are able to quantify how accuracy changes as damage increases (Figures \ref{TBI}, \ref{RelImportance}, \ref{Sparsified}, and \ref{Aging}) as well exactly which kinds of mistakes are being made (Figure \ref{ConfusionMatrices}). We demonstrate that the effect on accuracy is highly variable and depends on which connections are randomly selected (Figure \ref{RelImportance}), providing intuition for why impairments are difficult to predict. 

As with any model, using CNNs as an abstraction for the brain comes with limitations. Biological neural networks have many complex features and constraints that are not reflected in CNNs, such as transmitting information through spike trains and utilizing feedback. One important difference between convolutional neural networks and human subjects 
is the latter's ability to infer significantly more information from the \textit{context} of an image. 
For instance, a patient classifying all objects depicted in  Fig. \ref{ExamplesPepperGroup} 
might, due to some form of meta-analysis, readily interpret them as a collection of many-colored peppers. Consequently, he could discard extraneous objects from a list of 
candidates (like ping-pong/tennis balls) even if their shape and color alone do 
not provide sufficient evidence for such dismissal. We would encourage the usage  
of images with non-sensical pairings of objects to circumvent this difficulty in diagnostic 
tests for cognitive deficits. 

An interesting avenue for future work would be retraining a CNN after damaging it. To be biologically relevant, damaged weights would need to remain damaged, which is non-trivial to implement. However, the greater challenge is choosing an appropriate update rule. A key aspect of CNNs is taking advantage of some form of back propagation with gradient descent to solve the training optimization problem in a reasonable amount of time. This optimization problem is not convex, but for practical purposes, we generally do not worry about choosing a local optimum \cite{LossSurfaces}. However, to retrain a CNN after removing important weights, we may need to compensate by significantly changing other weights, escaping a local optimum. Even if we are already in the correct local optimum, it could be difficult to choose a step size. Moving in the direction of the gradient should move towards a local optimum unless the step size is too big. On the other hand, it will not converge quickly if the step size is too small. In the case of retraining a pre-trained but damaged network, some weights may be already optimal but others may need significant change, creating a difficult balance problem for choosing a step size. 

In summary, we provide a platform for quantitatively and qualitatively studying the progression of focal axonal swellings in a neural network. We can provide insight into disorders which feature FAS, such as TBI, Alzheimer's, Parkinson's, and Multiple Sclerosis, linking damage at the cellular level to changes in cognitive behavior.

\section*{Acknowledgments}
B. Lusch would like to acknowledge fellowship support from the National Physical Science Consortium and National Security Agency. 

\bibliography{CNNpaper,pedrobib}

\end{document}